\documentclass[aps,prl,twocolumn,longbibliography,floatfix,10pt,superscriptaddress]{revtex4-2}
\usepackage[utf8]{inputenc}
\usepackage{mathrsfs}
\usepackage{natbib}
\usepackage{graphicx}
\usepackage{latexsym}
\usepackage{amssymb}
\usepackage{amsmath}
\usepackage{amsfonts}
\usepackage{gensymb}
\usepackage{bm}
\usepackage{hyperref}
\usepackage{braket}
\usepackage{physics}
\usepackage{bbold}
\usepackage[caption=false]{subfig}
\usepackage[dvipsnames]{xcolor}
\usepackage[normalem]{ulem}
\usepackage{siunitx}
\usepackage{mhchem}
\usepackage{multirow}
\usepackage{soul}

\usepackage{slashed}

\definecolor{myforestgreen}{RGB}{34,139,34}

\newcommand{\PRLsec}[1]{\emph{#1---}}

\newcommand{\supplementarysection}{%
  \setcounter{figure}{0}
  \let\oldthefigure\thefigure
  \renewcommand{\thefigure}{S\oldthefigure}
  \setcounter{section}{0}
  \let\oldthesection\thesection
  \renewcommand{\thesection}{S\oldthesection}
  \setcounter{equation}{0}
  \let\oldtheequation\theequation
  \renewcommand{\theequation}{S\oldtheequation}
  \setcounter{table}{0}
  \let\oldthetable\thetable
  \renewcommand{\thetable}{S\oldthetable}
}

\newenvironment{dfn}{{\vspace*{1ex} \noindent \bf Definition }}{\vspace*{1ex}}

	\newcommand{\beq}{\begin{eqnarray}}
	\newcommand{\eeq}{\end{eqnarray}}
	\newcommand{\bea}{\begin{eqnarray}\begin{aligned}}
	\newcommand{\eea}{\end{aligned}\end{eqnarray}}

\begin{document}

\title{Anyon superfluid in trilayer quantum Hall systems}

\author{Taige Wang}
\affiliation{Department of Physics, University of California, Berkeley, CA 94720, USA \looseness=-2}
\affiliation{Kavli Institute for Theoretical Physics, University of California, Santa Barbara, CA 93106, USA \looseness=-2}
\author{Ya-Hui Zhang}
\affiliation{Department of Physics and Astronomy, Johns Hopkins University, Baltimore, Maryland 21218, USA \looseness=-2}

\begin{abstract}

Intertwining intrinsic topological order with gapless collective modes remains a central challenge in many-body physics.  We show that a quantum-Hall \emph{trilayer} at $\nu_{1}=\nu_{2}=\nu_{3}= \frac13$, tuned solely by the inter-layer spacing $d$, realizes this goal.  Large-scale density-matrix renormalization group (DMRG) calculations and a Chern-Simons field theory analysis reveal an intermediate ``anyon-exciton condensate'' separating the familiar $\nu_{\mathrm{tot}}=1$ exciton condensate ($d \to 0$) from three decoupled Laughlin liquids ($d \to \infty$).  In this phase, neutral bi-excitons condense while a $\nu=\frac23$ Laughlin topological order survives, yielding a Goldstone mode coexisting with fractionalized anyons.  A Ginzburg-Landau analysis maps out the finite-temperature phase diagram. The anyon-exciton condensate can be experimentally verified through  a vanishing double-counter-flow resistance and a fractional layer-resolved Hall resistivity \(R_{xy}=\frac{5}{2} h/e^{2}\), both within reach of existing high-mobility trilayer devices.
\end{abstract}

\maketitle

\PRLsec{Introduction} Topological order is usually protected by a bulk gap that endows anyons with finite energy and a ground state with degeneracy. Theory, however, has broadened the concept to \emph{gapless} topological order~\cite{SenthilFisher2000PRB,BondersonNayak2013PRB,Thorngren2021PRB,WenPotter2023PRB}. The most accessible member of this family would couple intrinsic topological order to a conventional \emph{superfluid}, whose Goldstone mode coexists with gapped anyons. Achieving such coexistence is non-trivial: the trivial boson that condenses in an ordinary superfluid typically lies high in energy inside a topological phase. A more promising route is to condense a composite anyon, echoing the “anyon superconductivity’’ proposals~\cite{Laughlin1988PRL,Laughlin1988Science,HalperinAnyonSC1990,FisherLeeAnyonSC1991}. Quantum-Hall multilayers are ideal for this purpose—each layer supplies anyons while neutral inter-layer excitons feel no magnetic field and can condense. In bilayers at $\nu=\frac13+\frac23$, this mechanism yields a \emph{trivial} exciton superfluid devoid of fractionalization~\cite{Zhang2023PRX}. We show that a richer state, in which superfluidity \emph{and} topological order coexist, naturally emerges in a quantum-Hall \emph{trilayer} at total filling $\nu_{\mathrm{tot}}=1$.

We consider three Coulomb-coupled layers separated by insulating barriers that suppress tunnelling, so each layer conserves charge independently. With a single Landau level per layer, the only tuning knob is the ratio between the inter-layer spacing $d$ and the magnetic length $ l_B$. Bilayers at $\nu_{\mathrm{tot}}=1$ are well understood, exhibiting an exciton superfluid for small $d/ l_B$~\cite{Moon1995PRB,Spielman2000PRL,Kellogg2004PRL,Tutuc2004PRL,Nandi2012Nature,Li2017NatPhys,Liu2017NatPhys,Liu2022Science}. By contrast, systems with $N\ge3$ remain largely unexplored. Focusing on the simplest case $\nu_1=\nu_2=\nu_3=\frac13$, we uncover an anyon-exciton condensate that retains intrinsic topological order—providing the sought-after coexistence in the cleanest possible setting.

In the $d/l_B \to 0$ limit, strong inter-layer Coulomb interaction locks the three layers together, breaks two independent $\mathrm{U}(1)$ symmetries, and produces a superfluid that generalizes the familiar $\nu_{\text{tot}} = 1$ bilayer condensate.  At the opposite extreme $d/l_B  \to \infty$, the layers decouple into three independent $\nu = \frac13$ Laughlin liquids. Combining field-theoretic analysis with large-scale DMRG on wide cylinders, we show that these limits are separated by a broad \emph{anyon-exciton condensate}: bi-exciton condensation coexists with Laughlin topological order of opposite chirality. The continuous transitions out of this phase belong to the XY$^{\ast}$ universality class~\cite{Zhang2023PRX}.

The phase admits an intuitive anyon-condensation picture.  In the decoupled limit each layer hosts Laughlin quasiparticles of charge $q=\pm e/3$.  As the spacing $d/l_B$ is reduced, inter-layer Coulomb interaction binds these anyons into excitons.  A naive choice $(q_1,q_2,q_3)=(+e/3,-e/3,0)$ remains anyonic and therefore cannot condense.  The simplest \emph{bosonic} composite $\Delta_{+}$ instead draws one anyon from each layer, carrying $(+e/3,-2e/3,+e/3)$. Because its net charge vanishes, $\Delta_{+}$ is blind to the external magnetic field and can condense. This condensation yields two crutial transport signatures: (i) a dissipationless counter-flow supercurrent in the $A_{+}$ channel and (ii) a \emph{fractionally} quantized Hall resistivity $R_{xy}=\frac{5}{2} \frac{h}{e^2}$  within the first layer.

In the \(d/l_B \to 0\) limit, we can analyze the finite temperature $T$ transition through the standard Ginzburg-Landau theory, which shows two separate Berezinskii-Kosterlitz-Thouless transitions. There is an intermediate tempeature regime with only a vestigial exciton condensation. Combining this finite temperature analysis with our zero-temperature results yields a complete \((d,T)\) phase diagram. High-mobility trilayer devices already exist~\cite{Jo1992PRB,Zeng2021Thesis}, putting our theoretical predictions within experimental reach in the near future.  
Our results provide a concrete blueprint for detecting anyon superfluidity that coexists with topological order and, in turn, for probing gapless topological phases.

\begin{figure}[t]
  \centering
	\includegraphics[width=0.9\columnwidth]{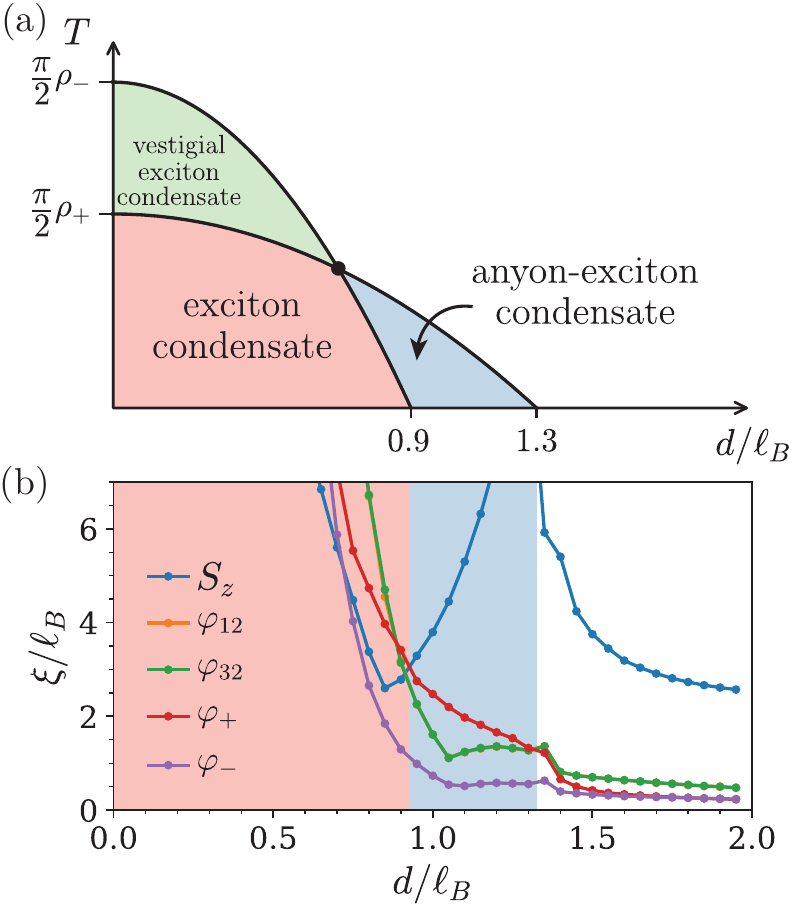}
	\caption{(a) Finite-temperature phase diagram of the trilayer as a function of inter-layer spacing \(d/ l_{B}\). At zero temperature three distinct ground states occur: a conventional exciton superfluid that breaks both inter-layer \(U(1)\) symmetries (red), an anyon–exciton condensate that breaks only \(U(1)_{+}\) symmetry while retaining topological order (blue), and three decoupled \(\nu=1/3\) Laughlin liquids (white). Thermal fluctuations disorder \(U(1)_{+}\) before \(U(1)_{-}\), resulting an intermediate vestigial exciton condensate (green). The two critical lines meet at a multicritical point. 
(b) Infinite-cylinder DMRG correlation lengths \(\xi_{\mathbf Q}\) versus \(d/ l_{B}\) in the five charge sectors defined in the text. Shaded regions reproduce the zero-temperature phases from panel (a). \(\xi_{\theta_1} = \xi_{\theta_2}\) due to layer-swap symmetry. Here we have used cylinder circumference \(L_y= 14 l_B\) and bond dimension \(\chi = 10000\).}
\label{fig:pd}
\end{figure}

\PRLsec{Quantum Hall trilayers} We begin by organizing the microscopic symmetries that constrain the infrared theory of a \(\nu=\frac13+\frac13+\frac13\) trilayer.  
Because each layer separately conserves charge, the system hosts three independent \(U(1)\) symmetries.  Writing the layer charges as \(Q_{\alpha}\), the combinations  
\(Q_c = Q_{1}+Q_{2}+Q_{3}\),  
\(\Delta Q_{12}=Q_{1}-Q_{2}\), and  
\(\Delta Q_{32}=Q_{3}-Q_{2}\)  
generate the total-charge symmetry \(U(1)_{c}\) and two inter-layer symmetries \(U(1)_{12}\) and \(U(1)_{32}\).  
A discrete layer exchange \(\mathbb{Z}_{2}:1\leftrightarrow3\) swaps the outer layers, sending \(\Delta Q_{12}\leftrightarrow\Delta Q_{32}\). 

At zero spacing \(d=0\) these abelian symmetries enhance to an \(SU(3)\) flavour group, and the ground state is an \(SU(3)\) quantum-Hall ferromagnet.  
A small but finite spacing selects the easy-plane coherent state
\begin{equation}
|\Psi\rangle = \prod_{m}
\bigl(
  e^{i \varphi_{12}} c^{\dagger}_{1m} +
                   c^{\dagger}_{2m}   +
  e^{i \varphi_{32}} c^{\dagger}_{3m}
\bigr)|0\rangle,
\end{equation}
where \(m\) labels Landau-gauge orbitals.  
The phases \( \varphi_{12}\) and \( \varphi_{32}\) represent \emph{independent} exciton condensates that break \(U(1)_{12}\) and \(U(1)_{32}\). This state is the direct trilayer analogue of the bilayer exciton superfluid first analysed by Moon \textit{et al.}~\cite{Moon1995PRB}.  
In the opposite limit \(d\to\infty\) the layers decouple into three \(\nu=1/3\) Laughlin liquids, restoring both inter-layer \(U(1)\)'s.

Between these extremes our numerics uncovers an \emph{intermediate} regime in which the composite order parameter \( e^{i \varphi_{12}} e^{i \varphi_{32}}\) attains long-range coherence while the single excitons \( e^{i \varphi_{12}}\) and \( e^{i \varphi_{32}}\) remain short-ranged.  
Such ``paired'' condensation echoes charge-\(4e\) or charge-\(6e\) superconductors \cite{Almoalem2022NatPhys,Iyuha2024LP,Iguchi2023Science,Ge2024PRX}.

To expose the relevant symmetry, we introduce the combinations
\begin{equation}
     \varphi_{+}= \varphi_{12} + \varphi_{32},\qquad
 \varphi_{-}= \varphi_{12} - \varphi_{32},
\end{equation}
which characterize the rotated \(U(1)\) symmetries \(U(1)_{+}\) and \(U(1)_{-}\). Under the swap \(\mathbb{Z}_{2}\) one has \( \varphi_{\pm}\to\pm \varphi_{\pm}\). Thus condensing \(e^{i \varphi_{\pm}}\) breaks \(U(1)_{\pm}\) while preserving both \(U(1)_{\mp}\) and the layer swap. We can further define the corresponding background gauge fields to be
\begin{equation}
    A_c = \frac13 \sum_{\alpha} A_{\alpha}, \; A_+ = \frac{A_1-2A_2+A_3}2, \; A_- = \frac{A_1-A_3}2
\end{equation}
where $A_{\alpha}$ is the background gauge field on the ${\alpha}$th layer. Our numerics shows that the intermediate phase in which \(e^{i \varphi_{+}}\) is condensed, potentially resulting an exciton superfluid that coexists with Laughlin topological order, as elucidated by the Chern–Simons theory that follows.

\PRLsec{Numerical calculation} We simulate three parallel two-dimensional electron layers at total filling $\nu_{1}=\nu_{2}=\nu_{3}= \frac13$ separated by a distance \(d\).  Prior to Landau-level projection the Hamiltonian reads
\begin{equation}
\label{eq:H_def}
H=\frac12\sum_{\alpha,\beta=1}^{3}\int_{\mathbf r}\int_{\mathbf r'}\hat\rho_{\alpha}(\mathbf r)V_{\alpha\beta}(\mathbf r-\mathbf r')\hat\rho_{\beta}(\mathbf r'),
\end{equation}
with \(\hat\rho_{\alpha}(\mathbf r)\) is the bare density in layer \(\alpha\), and \(V_{\alpha\beta}(\mathbf q)=e^{-q|\alpha-\beta|d}e^{2}/\epsilon q\) is the bare Coulomb repulsion. A metallic gate is placed \(D=20 l_{B}\) away to improve convergence.  After projecting to the lowest Landau level on an infinite cylinder of circumference \(L_{y}=14 l_{B}\), we solve the Hamiltonian with infinite density matrix renormalization group (iDMRG) upto bond dimension \(\chi =  10000\). We conserve the individual layer charges \(Q_{1},Q_{2},Q_{3}\) and the total \(k_{y}\) momentum  \cite{ExactMPS,TopoChara}.

In quasi-1D numerics the \(U(1)\) symmetries cannot break spontaneously.  We therefore keep all \(U(1)\)’s intact and diagnose their fate in two dimensions by examining fluctuations in the corresponding charge sectors. For each spacing \(d/l_B\) we diagonalize the matrix product state (MPS) transfer matrix and extract the leading correlation length \(\xi_{\hat O}\) for operator $\hat O$ in five charge sectors:
\begin{equation}
\begin{gathered}
S_{z}:(0,0,0),\quad
\varphi_{-}:(e,0,-e),\\
\varphi_{12}:(e,-e,0),\;
\varphi_{32}:(0,-e,e),\;
\varphi_{+}:(e,-2e,e).
\end{gathered}
\end{equation}
The \(\varphi_{12,32}\) channels probe nearest-layer excitons; \(\varphi_{\pm}\) probe the composite order.  Layer-swap symmetry enforces \(\xi_{\varphi_{12}}=\xi_{\varphi_{32}}\).  Fig.~\ref{fig:pd} (b) plots all correlation lengths versus \(d/ l_{B}\).

For \(d/ l_{B}<0.9\) the longest correlation lengths lie in the \(\varphi_{12,32}\) sectors, indicating that both inter-layer excitons condense. This is a conventional $\nu_{\mathrm{tot}} = 1$ exciton condensate which connects adiabatically to the \(SU(3)\) quantum Hall ferromagnet in the \(d=0\) limit.  In the intermediate window \(0.9<d/ l_{B}<1.3\) the \(\varphi_{+}\) channel overtakes \(\varphi_{12,32}\) and \(\xi_{S_{z}}\) becomes largest, signaling selective condensation of a composite order parameter $e^{i\varphi_+}\sim c^\dagger_1 c^\dagger_3 c_2 c_2$. This phase only breaks the \(U(1)_{+}\) symmetry. For \(d/ l_{B}>1.3\) all inter-layer correlation lengths become short, consistent with three decoupled \(\nu=1/3\) Laughlin liquids. Though direct eight-fermion correlators of $c_1^\dagger c_3^\dagger c_2 c_2$ are impractical, the hierarchy of  the correlation length \(\xi_{\hat O}\) provides a clean diagnostic of the zero-temperature numerical phase diagram in Fig.~\ref{fig:pd} (a).

At first sight it is puzzling why the two elementary excitons $c^\dagger_1 c_2$ and $c_3^\dagger c_2$ want to bind together despite their dipolar repulsion.
We propose a kinetic mechanism that parallels the idea of anyon superconductivity \cite{Laughlin1988PRL,HalperinAnyonSC1990,FisherLeeAnyonSC1991}. Starting from the decoupled phase in the large $d$ limit, we have simple anyon-exciton with charge $(e/3,-e/3,0)$ and $(0,-e/3,e/3)$. However, each of them has a fractional statistics, which frustrates its coherent motion. In contrast, a par of them with charge $(e/3,-2e/3,e/3)$ is a neutral boson and can condense. As we show later, the condensation of this fractional boson leads to a superfluid phase with  $\langle e^{i\varphi_+} \rangle \neq 0$ order, but with a residual coexisting topological order. 

Note that the fractional anyon-exciton $\Delta_+$ with charge $(e/3,-2e/3,e/3)$  is not gauge invariant and its correlation function is not detectable in numerical and experimental measurements. In our DMRG calculation, we can only get access to the gauge invariant operator $e^{i\varphi_+}\sim \Delta_+^3$. Strictly speaking, we do not have direct evidence that the intermediate phase is formed through condensation of $\Delta_+$ instead of $\Delta_+^3$.  If only $e^{i\varphi_+}$ is condensed, then the resulting superfluid phase has exactly the same  topological order as the decoupled Laughlin state. On the other hand, in our construction below, the condensation of $\Delta_+$ gives the same symmetry breaking order, but also partially confines some anyons and leads to a different topological order. These two scenarios can only be distinguished through the number of anyons survived, which is quite challenging for numerics. Energetically, however, repulsive interactions make it costly to bind three $\Delta_{+}$ excitations before condensation, rendering direct $\Delta_{+}$ condensation the more natural scenario. We therefore focus on that case in what follows.

\PRLsec{Anyon-exciton condensate} In the \(d\gg \ell_B\) limit, the long-wavelength dynamics is captured by the Chern–Simons Lagrangian
\begin{equation}
\label{eq:L0}
\mathcal L_{0}= -\frac{1}{4\pi} a^{\mathsf T}K da + \frac{1}{2\pi} a^{\mathsf T} Q dA ,
\end{equation}
with \(K=\mathrm{diag}(3,3,3)\), \(Q=\mathrm{diag}(1,1,1)\), \(a = (a_1,a_2,a_3)^T\) are three dynamical gauge fields for each layer. Any excitation is specified by an integer vector \(l=( l_1, l_2, l_3)\) that assigns layer charge \(Q_\alpha =  l_\alpha e/3\) and carries exchange phase \(\theta/2\pi = l^{\mathsf T}K^{-1}l\).

Almost all inter-layer excitons are anyons with \(\theta\neq0\).  The lone self-boson is the ``symmetric'' exciton \(\Delta_+\) with charge vector
\begin{equation}
l_{+} = (1,-2,1),
\end{equation}
for which \(l_{+}^{\mathsf T}K^{-1}l_{+}=2\) and hence \(\theta_{+}=2\pi\).  Self-bosonic statistics is necessary but not sufficient for condensation.  Introducing a matter field \(\tilde \varphi\) that carries charge \(l_{+}\),
\begin{equation}
\mathcal L = \mathcal L_{0} + \bigl|(\partial_\mu - il_{+}^{\mathsf T}a_\mu)\tilde \varphi\bigr|^{2} + m|\tilde \varphi|^{2},
\end{equation}
From variation of $a_0$ one finds that $\frac{1}{2\pi} d \bigl(l_{+}^{\mathsf T}a\bigr)=2 n_{\varphi}$ so $\tilde \varphi$ feels effectively two flux and can not simply condense.

Following Laughlin’s anyon-superconductivity prescription~\cite {Laughlin1988PRL,HalperinAnyonSC1990,FisherLeeAnyonSC1991}, we can attach to fluxes to the boson and then condense the composite boson.
\begin{equation}
\label{eq:Lbind}
\mathcal L_{\mathrm{critical}} = \mathcal L_{0} + \bigl|(\partial_\mu - il_{+}^{\mathsf T}a_\mu-i \tilde \beta_\mu) \varphi\bigr|^{2} + m| \varphi|^{2}+\frac{1}{8\pi} \tilde \beta d \tilde \beta
\end{equation}
where $\tilde \beta$ is introduced to implement the flux attachment to get the composite boson $\varphi$. The above theory describes the transition from the decoupled phase to the anyon superfluid phase by tuning $m$. In the side with $\varphi$ condensed, we can get an effective theory by introducing a dual gauge field $\beta$ for the condensation of $\varphi$:

\begin{equation}
    \mathcal L=\mathcal L_0-\frac{1}{2\pi} (l_+^T a+\tilde \beta) d \beta+\frac{1}{8\pi} \tilde \beta d \tilde \beta
\end{equation}
We can integrate $\tilde \beta$ and get a simplified theory
\begin{equation}
    \mathcal L=\mathcal L_0-\frac{2}{4\pi} \beta d \beta-\frac{1}{2\pi} (l_+^{\mathsf T} a) d \beta
\end{equation}
It is intuitive to think that we put the boson $\tilde \varphi$ in a $\nu=\frac{1}{2}$ bosonic Laughlin state. Finally in the anyon superfluid phase, we have a new $K$ matrix
\begin{equation}
\tilde K=
\begin{pmatrix}
3&0&0&1\\
0&3&0&-2\\
0&0&3&1\\
1&-2&1&2
\end{pmatrix}
\end{equation}
\(\tilde K\) has one null vector and three positive eigenvalues: the null direction is the Goldstone mode of broken \(U(1)_{+}\); the positive block retains the chiral central charge \(c=3\) of the parent Laughlin liquids.  The phase is therefore an \emph{anyon-exciton condensate}: topological order coexisting with a single neutral superfluid mode.

Because any quasiparticle that braids non-trivially with the condensed boson \(\Delta_{+}\) is confined, the deconfined spectrum collapses to three linearly independent vectors,
\begin{equation}
l_{0}=(0,0,0),\qquad
l_{1}=(0,1,2),\qquad
l_{2}=(0,2,1),
\end{equation}
defined modulo attachment of \(l_{+}\) or an electron \((1,1,1)\). The super-section sectors therefore form a \(\mathbb Z_{3}\) group, implying a three-fold ground‐state degeneracy on the torus.

To expose the residual topological data we perform a unimodular change of variables that separates the Goldstone mode \(\tilde K \xrightarrow{W_{1}} K' \) (see End Matter for detailed derivation), where
\begin{equation}
K'= (0) \oplus
\begin{pmatrix}
3 & 0 & 2\\
0 & 3 & 1\\
2 & 1 & 2
\end{pmatrix}
\end{equation}
The empty first row and column of \(K'\) corresponds to the Goldstone mode \(\gamma\), and \(A_+\) is Higgsed out. 

After decoupling the Goldtsone mode and attaching a neutral integer quantum Hall (QH) layer \((-1)\), a second
transformation \(W_{2}\) diagonalizes the \(K\) matrix, \(K' \oplus(-1) \xrightarrow{W_{2}} K'' = (-3)\oplus\mathbf 1_{3} \). 
Therefore, the anyon-exciton condensate is stably equivalent to the Laughlin state at
\(\nu=-1/3\) plus four integer-QH layers.

After restroing the coupling to the background gauge field, we arrive at
\begin{equation} \label{eq:action}
\begin{aligned}
    \mathcal L=& - \frac{1}{4\pi} \tilde a_-^{\mathsf T} \begin{pmatrix}
        1 & 0\\
        0 & -3
    \end{pmatrix} d \tilde a_- + \frac{1}{2\pi} \tilde a_-^{\mathsf T} d \tilde A_-\\
    &-\frac{1}{4\pi} \tilde a_c d \tilde a_c +\frac{1}{2\pi} \tilde a_c d \tilde A_c -\frac{1}{2\pi} \gamma  d \tilde A_+ + 4 \mathrm{CS}_g.
\end{aligned}
\end{equation}
where $\gamma$ is an internal gauge field which represents the Goldstone mode, and $\mathrm{CS}_g$ is the gravitational Chern-Simons term. The term $- \gamma d \tilde A_+$ gives the superfluid phase in the $\tilde A_+=A_1-2A_2+A_3$ channel. The residue topological order is therefore a $\nu=2/3$ Laughlin state in the $\tilde A_-=A_2-A_3$ channel (described by a two-component gauge field $\tilde a_-$) and an integer-QH state in the $\tilde A_c=A_2$ channel, accompanied by two neutral integer-QH layers responsible for $\mathrm{CS}_g$.

We can integrate out the gauge fields and obtain the response theory
\begin{equation} \label{eq:response}
\mathcal L_{\mathrm{res}}=\frac{1}{4\pi} A_cd  A_c+\frac{1}{6\pi} A_{-}d  A_{-}
+\frac{\rho_{s}}{2}\bigl(\partial_{\mu}\varphi_{+}-A_{+,\mu}\bigr)^{2}.
\end{equation}
Since $\gamma$ Higgesed the $A_+$ field, we have $d A_+ = 0$, and consequently $d A_- = d \tilde A_-$ and $d A_c=d \tilde A_c$.  The superfluid component makes the conductance tensor $\sigma_{xy}$ ill-defined, but the resistivity tensor $\rho_{yx}$ is well defined and can be found to be
\begin{equation} \label{eq:rhoxy}
    \rho_{yx}=\frac{h}{e^2} \begin{pmatrix}
\frac{5}{2} & 1 & -\frac{1}{2} \\
1 & 1 & 1 \\
-\frac{1}{2} & 1 & \frac{5}{2}
\end{pmatrix}
\end{equation}
Note that there is fractional  Hall resistivity in contrast to the conventional exciton superfluid phase. Driving current $I_{1;x}=I$ in the first layer leads to a Hall voltage drop $V_{1;y}=\frac{5}{2} \frac{h}{e^2}I$ in the first layer and $V_{3;y}=-\frac{1}{2} \frac{h}{e^2}I$ in the third layer. For the small $d/ l_B$ exciton condensate phase, all entries of $\rho_{yx}$ should be $1$; for the decoupled state at larger $d/ l_B$, $\rho_{yx}$ is proportional to the identity matrix with coefficient $3 h/e^2$. Hence simple layer-resolved measurement of Hall resistivity can distinguish the intermediate anyon-superfluid phase from the other two phases in the small and large $d/l_B$ limit.

\PRLsec{Finite–temperature phase diagram} For spacings \(d\ll l_{B}\) all anyonic quasiparticles are gapped, so the low–energy physics reduces to the two compact inter-layer phases \(\varphi_{12},\varphi_{32}\). Amplitude and out-of-plane fluctuations remain massive and can be neglected. Coulomb exchange penalizes spatial variations of the phase difference between \textit{any} pair of layers.  Gradients of the adjacent phases \(\varphi_{12}\) and \(\varphi_{32}\) cost the nearest–layer stiffness \(\rho_{\mathrm{N}}\), whereas gradient of the outer–layer phase difference \(\varphi_{12}-\varphi_{32}\) incur additional stiffness \(\rho_{\mathrm{O}}\).  Collecting the two contributions yields the rotationally invariant Ginzburg–Landau functional
\begin{equation}
\label{eq:GL}
E=\frac12\int d^{2}\mathbf r
\Bigl[
\rho_{\mathrm{N}}\bigl(|\nabla\varphi_{12}|^{2}+|\nabla\varphi_{32}|^{2}\bigr)
+\rho_{\mathrm{O}}|\nabla(\varphi_{12}-\varphi_{32})|^{2}
\Bigr],
\end{equation}
where \(\rho_{\mathrm{N}}\) and \(\rho_{\mathrm{O}}\) are, respectively, the nearest-layer and outer-layer stiffnesses. 
Equivalently,  
\begin{equation}
E=\frac12\int d^{2}\mathbf r\nabla\varphi_{\alpha}
\rho_{\alpha\beta}
\nabla\varphi_{\beta},\quad
\rho_{\alpha\beta}=
\begin{pmatrix}
\rho_{\mathrm{N}}+\rho_{\mathrm{O}} & -\rho_{\mathrm{O}}\\
-\rho_{\mathrm{O}} & \rho_{\mathrm{N}}+\rho_{\mathrm{O}}
\end{pmatrix}.
\end{equation}

The eigen‐phases \(\varphi_{\pm}=\varphi_{12} \pm \varphi_{32}\) diagonalize the stiffness matrix in Eq.\eqref{eq:GL}, giving  
\begin{equation}
\rho_{+}=\rho_{\mathrm{N}},\qquad \rho_{-}=\rho_{\mathrm{N}}+2\rho_{\mathrm{O}}.
\end{equation}
Because \(\rho_{\mathrm{O}}>0\) for all \(d\ll l_{B}\) we always have \(\rho_{-}>\rho_{+}\).  Each compact \(U(1)\) field disorders via its own Berezinskii-Kosterlitz-Thouless (BKT) transition at temperature
\begin{equation}
T_{\mathrm{BKT}}^{\pm}=\frac{\pi}{2}\rho_{\pm},
\end{equation}
Because \(T_{\mathrm{BKT}}^{-}>T_{\mathrm{BKT}}^{+}\), the two compact phases lose coherence at parametrically separated temperatures. Heating the  exciton superfluid first drives a BKT transition at \(T_{\mathrm{BKT}}^{+}\), which disorders \(\varphi_{+}\) but leaves \(\varphi_{-}\) stiff.  The system thereby enters an \emph{intermediate} phase that only breaks the \(U(1)_{-}\) symmetry.  Although the corresponding exciton \(\Delta_{-}\) involves only the two outer layers, its stiffness still enjoys nearest layer coupling \(\rho_{\mathrm{N}}\): any gradient in \(\varphi_{-}\) necessarily perturbs charge balance with the middle layer.

Fig.~\ref{fig:pd}(a) collects the full \(d-T\) phase diagram from numerical calculation and Ginzburg–Landau analysis. At small spacing and low temperature (red) both \(U(1)\) symmetries are spontaneously broken, realizing the conventional exciton superfluid. Increasing \(d\) drives an \(\mathrm{XY}^{\ast}\) quantum transition into the anyon-exciton condensate (blue), which breaks only \(U(1)_{+}\) and supports Laughlin topological order. Raising the temperature at fixed \(d\) first melts \(U(1)_{+}\) through an BKT transition at \(T_{\mathrm{BKT}}^{+}\), producing a vestigial order that breaks only \(U(1)_{-}\) (green), and finally all $U(1)$ symmetries are restored at a higher temperature \(T_{\mathrm{BKT}}^{-}\). We anticipate that the two critical lines merge at a single multicritical point (black dot).

\PRLsec{Experimental signatures}
A single Hall bar can unambiguously expose the anyon–exciton condensate phase.  
Driving double counter-flow currents $(+I,-2I,+I)$ excites the $U(1)_{+}$ channel; the resulting \emph{vanishing} longitudinal resistance evidences a dissipationless neutral supercurrent.  Meanwhile the usual counter-flow transport in the  $(+I,-I,0)$ channel behaves as an insulator in this intermediate phase, in contrast to a superfluid in the small $d/l_B$ phase.
Simultaneously, layer-resolved Hall resistivities\cite{Li2017NatPhys,Liu2017NatPhys}.  exhibit \emph{fractional} plateaus
\begin{equation}
    \rho_{yx}^{11}= \frac{5}{2}\,\frac{h}{e^{2}}, \qquad
\rho_{yx}^{13}= -\frac12\,\frac{h}{e^{2}},
\end{equation}
confirming the existence of fractionalization. 
A complementary probe monitors layer-resolved compressibilities~\cite{Liu2021PRL,Zeng2021Thesis,Zibrov2017Nature}: the $U(1)_+$ channel is compressible, while the $U(1)_{c}$ and $U(1)_{-}$ channels are incompressible.  
Finally, ordering of $U(1)_{+}$ should yield a pronounced negative Coulomb-drag peak~\cite{Hartman2018PRB,Nandi2012Nature}.  
Converging evidence from these measurements would establish the first gapless topological phase in two dimensions and open a pathway to Josephson-like devices between fractional states~\cite{Wang2024JJ}.

\PRLsec{Conclusion}
We predict that a charge-balanced trilayer at $\nu_{1}=\nu_{2}=\nu_{3}= \frac13$ realizes an \emph{anyon-exciton condensate} at intermediate layer separation: a phase where superfluidity  coexists with topological order. Starting from the decoupled Laughlin state in the large distance limit, condensation of a pair of excitons formed by anyons leads to a bi-exciton superfluid phase, with residual topological order equivalent to $\nu=\frac{2}{3}$ Laughlin state. We make predictions in Hall resistivity and counterflow transport, which can be tested in near-term experiments. Our work shows that quantum Hall trilayer may host richer physics than the well-studied quantum Hall bilayer. We anticipate more discoveries to be made at other fillings. \\

\textbf{Acknowledgments.} We especially thank Michael P. Zaletel, Roger S. K. Mong, and Frank Pollmann for developing the iDMRG codes for multicomponent quantum Hall systems. We thank Yihang Zeng, Tianle Wang and Darius Zhengyan Shi for helpful discussions. We thank Zhaoyu Han, Zhihuan Dong, Michael P. Zaletel, and Ashvin Vishwanath for collaboration on a related project. TW is supported by the Simons Collaboration on Ultra-Quantum Matter, which is a grant from the Simons Foundation (Grant No. 1151944, MZ). TW is also supported by the Heising-Simons Foundation, the Simons Foundation, and NSF grant No. PHY-2309135 to the Kavli Institute for Theoretical Physics (KITP). YHZ is supported by the National Science Foundation under Grant No. DMR-2237031. This research uses the Lawrencium computational cluster provided by the Lawrence Berkeley National Laboratory (Supported by the U.S. Department of Energy, Office of Basic Energy Sciences under Contract No. DE-AC02-05-CH11231).

\begin{center}
    \textbf{End Matter}
\end{center}

In the End Matter we detail the derivation of the underlying topological quantum field theory (TQFT).  
In general, any Abelian topological order is characterized by a $K$-matrix and a charge matrix $Q$,
\begin{equation}
    \mathcal L_{0}= -\frac{1}{4\pi} a^{\mathsf T} K  d a + \frac{1}{2\pi} a^{\mathsf T} Q  dA ,
\end{equation}
where $a=(a_1,\dots ,a_N)$ are $N$ dynamical gauge fields and  
$A=(A_1,\dots ,A_M)$ are $M$ background gauge fields.  
For a single background field ($M=1$) the matrix $Q$ reduces to the familiar charge vector $t$.  
Here there are three physical layers, so $M=3$.  
Three decoupled Laughlin states are therefore described by
\begin{equation}
    K=\begin{pmatrix}3&0&0\\ 0&3&0\\ 0&0&3\end{pmatrix},\quad 
    Q=\begin{pmatrix}1&0&0\\ 0&1&0\\ 0&0&1\end{pmatrix}.
\end{equation}

When the exciton $\Delta_+$ condenses into a bosonic $\nu=\frac12$ Laughlin state, the system enters an anyon-exciton condensate.  
Introducing a fourth dynamical gauge field $\beta$ to describe this bosonic Laughlin state, the theory becomes
\begin{equation}
\tilde K=\begin{pmatrix}
3&0&0&1\\
0&3&0&-2\\
0&0&3&1\\
1&-2&1&2
\end{pmatrix},\quad 
\tilde Q=\begin{pmatrix}1&0&0\\ 0&1&0\\ 0&0&1\\ 0&0&0\end{pmatrix}.
\end{equation}
where $\tilde a=(a_1,a_2,a_3,\beta)$.

To expose the residual topological order we perform unimodular transformations $W\in\mathrm{SL}(N,\mathbb Z)$,
\begin{equation}
    K \to  W^{\mathsf T} K W,\qquad Q \to  W^{\mathsf T} Q .
\end{equation}
where $W$ is an integer matrix with $\det W = 1$.
Applying
\begin{equation}
    W_1=\begin{pmatrix}-1&0&0&0\\ 2&-1&0&0\\ -1&0&1&0\\ 3&0&0&1\end{pmatrix},
\end{equation}
we obtain
\begin{equation}
\tilde{K} \to K'=\begin{pmatrix}
0&0&0&0\\
0&3&0&2\\
0&0&3&1\\
0&2&1&2
\end{pmatrix},\quad 
\tilde{Q} \to Q'=\begin{pmatrix}-1&2&-1\\ 0&-1&0\\ 0&0&1\\ 0&0&0\end{pmatrix}.
\end{equation}
The vanishing first row and column identify the Goldstone mode $\gamma$, so $a' = (\gamma,a_1',a_2',a_3')$. From the first column of $Q'$, we see that \(\tilde A_+ = A_1 - 2A_2 + A_3\) is Higgsed out by the Goldstone mode, i.e. \(- \gamma d A_+\). 

After eliminating $\gamma$ and appending a neutral integer-QH layer, we have
\begin{equation}
\tilde K' =\begin{pmatrix}
3&0&2&0\\
0&3&1&0\\
2&1&2&0\\
0&0&0&-1
\end{pmatrix},\quad 
\tilde Q'=\begin{pmatrix}0&-1&0\\ 0&0&1\\ 0&0&0\\ 0&0&0\end{pmatrix}.
\end{equation}
A second unimodular transformation
\begin{equation}
    W_2=\begin{pmatrix}-1&-1&0&-1\\ -1&0&0&-1\\ 0&1&1&1\\ 3&0&-1&1\end{pmatrix}
\end{equation}
yields
\begin{equation}
\tilde K' \to K''=\begin{pmatrix}-3&0&0&0\\ 0&1&0&0\\ 0&0&1&0\\ 0&0&0&1\end{pmatrix},\quad 
\tilde Q' \to Q''=\begin{pmatrix}0&1&-1\\ 0&1&0\\ 0&0&0\\ 0&1&-1\end{pmatrix}.
\end{equation}
The final form makes the physics transparent: the anyonic superfluid lives in the $\tilde A_+$ channel, coexisting with a $\nu=2/3$ Laughlin state in the $\tilde A_- = A_2-A_3$ channel (first and last row of $Q''$) and an integer QH state in the $\tilde A_c = A_2$ channel (second row of $Q''$).  
The appended neutral layer and the third row of $Q''$ constituent two neutral integer-QH states.  
The resulting effective action is given in Eq.~\eqref{eq:action}.

We define the layer-resolved Hall-resistivity tensor $\rho_{yx}$ via
\begin{equation}
    E_\alpha^{y}= \sum_{\beta=1}^{3} \rho_{yx}^{\alpha\beta}  I_\beta^{x},
    \qquad \alpha, \beta=1,2,3,
\end{equation}
where $E_\alpha^{x}$ is the in-plane electric field on layer $\alpha$ and $I_\beta^{y}$ is the transverse current flowing in layer $\beta$. 
To extract $\rho_{yx}$ from the response theory of Eq.~\eqref{eq:response}, we note that its three independent terms imply
\begin{equation}
    \begin{cases}
        \frac13\bigl(E_1^{y}+E_2^{y}+E_3^{y}\bigr)= I_1^{x}+I_2^{x}+I_3^{x},\\
        E_1^{y}-E_3^{y}=3\bigl(I_1^{x}-I_3^{x}\bigr),\\
        \mathbf E_1+\mathbf E_3 = 2\mathbf E_2,
    \end{cases}
\end{equation}
the last relation arising from Meissner effect.  
Together with layer-exchange $\mathbb Z_2$ symmetry and the intrinsic symmetry $\rho_{yx}^{\alpha\beta}=\rho_{yx}^{\beta\alpha}$, these constraints fix a unique tensor presented in Eq.~\eqref{eq:rhoxy}.

\bibliography{main}

\begin{thebibliography}{29}%
\makeatletter
\providecommand \@ifxundefined [1]{%
 \@ifx{#1\undefined}
}%
\providecommand \@ifnum [1]{%
 \ifnum #1\expandafter \@firstoftwo
 \else \expandafter \@secondoftwo
 \fi
}%
\providecommand \@ifx [1]{%
 \ifx #1\expandafter \@firstoftwo
 \else \expandafter \@secondoftwo
 \fi
}%
\providecommand \natexlab [1]{#1}%
\providecommand \enquote  [1]{``#1''}%
\providecommand \bibnamefont  [1]{#1}%
\providecommand \bibfnamefont [1]{#1}%
\providecommand \citenamefont [1]{#1}%
\providecommand \href@noop [0]{\@secondoftwo}%
\providecommand \href [0]{\begingroup \@sanitize@url \@href}%
\providecommand \@href[1]{\@@startlink{#1}\@@href}%
\providecommand \@@href[1]{\endgroup#1\@@endlink}%
\providecommand \@sanitize@url [0]{\catcode `\\12\catcode `\$12\catcode `\&12\catcode `\#12\catcode `\^12\catcode `\_12\catcode `\%12\relax}%
\providecommand \@@startlink[1]{}%
\providecommand \@@endlink[0]{}%
\providecommand \url  [0]{\begingroup\@sanitize@url \@url }%
\providecommand \@url [1]{\endgroup\@href {#1}{\urlprefix }}%
\providecommand \urlprefix  [0]{URL }%
\providecommand \Eprint [0]{\href }%
\providecommand \doibase [0]{https://doi.org/}%
\providecommand \selectlanguage [0]{\@gobble}%
\providecommand \bibinfo  [0]{\@secondoftwo}%
\providecommand \bibfield  [0]{\@secondoftwo}%
\providecommand \translation [1]{[#1]}%
\providecommand \BibitemOpen [0]{}%
\providecommand \bibitemStop [0]{}%
\providecommand \bibitemNoStop [0]{.\EOS\space}%
\providecommand \EOS [0]{\spacefactor3000\relax}%
\providecommand \BibitemShut  [1]{\csname bibitem#1\endcsname}%
\let\auto@bib@innerbib\@empty
\bibitem [{\citenamefont {Senthil}\ and\ \citenamefont {Fisher}(2000)}]{SenthilFisher2000PRB}%
  \BibitemOpen
  \bibfield  {author} {\bibinfo {author} {\bibfnamefont {T.}~\bibnamefont {Senthil}}\ and\ \bibinfo {author} {\bibfnamefont {M.~P.~A.}\ \bibnamefont {Fisher}},\ }\bibfield  {title} {\bibinfo {title} {${Z}_{2}$ gauge theory of electron fractionalization in strongly correlated systems},\ }\href {https://doi.org/10.1103/PhysRevB.62.7850} {\bibfield  {journal} {\bibinfo  {journal} {Phys. Rev. B}\ }\textbf {\bibinfo {volume} {62}},\ \bibinfo {pages} {7850} (\bibinfo {year} {2000})}\BibitemShut {NoStop}%
\bibitem [{\citenamefont {Bonderson}\ and\ \citenamefont {Nayak}(2013)}]{BondersonNayak2013PRB}%
  \BibitemOpen
  \bibfield  {author} {\bibinfo {author} {\bibfnamefont {P.}~\bibnamefont {Bonderson}}\ and\ \bibinfo {author} {\bibfnamefont {C.}~\bibnamefont {Nayak}},\ }\bibfield  {title} {\bibinfo {title} {Quasi-topological phases of matter and topological protection},\ }\href {https://doi.org/10.1103/PhysRevB.87.195451} {\bibfield  {journal} {\bibinfo  {journal} {Phys. Rev. B}\ }\textbf {\bibinfo {volume} {87}},\ \bibinfo {pages} {195451} (\bibinfo {year} {2013})}\BibitemShut {NoStop}%
\bibitem [{\citenamefont {Thorngren}\ \emph {et~al.}(2021)\citenamefont {Thorngren}, \citenamefont {Vishwanath},\ and\ \citenamefont {Verresen}}]{Thorngren2021PRB}%
  \BibitemOpen
  \bibfield  {author} {\bibinfo {author} {\bibfnamefont {R.}~\bibnamefont {Thorngren}}, \bibinfo {author} {\bibfnamefont {A.}~\bibnamefont {Vishwanath}},\ and\ \bibinfo {author} {\bibfnamefont {R.}~\bibnamefont {Verresen}},\ }\bibfield  {title} {\bibinfo {title} {Intrinsically gapless topological phases},\ }\href {https://doi.org/10.1103/PhysRevB.104.075132} {\bibfield  {journal} {\bibinfo  {journal} {Phys. Rev. B}\ }\textbf {\bibinfo {volume} {104}},\ \bibinfo {pages} {075132} (\bibinfo {year} {2021})}\BibitemShut {NoStop}%
\bibitem [{\citenamefont {Wen}\ and\ \citenamefont {Potter}(2023)}]{WenPotter2023PRB}%
  \BibitemOpen
  \bibfield  {author} {\bibinfo {author} {\bibfnamefont {R.}~\bibnamefont {Wen}}\ and\ \bibinfo {author} {\bibfnamefont {A.~C.}\ \bibnamefont {Potter}},\ }\bibfield  {title} {\bibinfo {title} {Bulk–boundary correspondence for intrinsically gapless symmetry-protected topological phases from group cohomology},\ }\href {https://doi.org/10.1103/PhysRevB.107.245127} {\bibfield  {journal} {\bibinfo  {journal} {Phys. Rev. B}\ }\textbf {\bibinfo {volume} {107}},\ \bibinfo {pages} {245127} (\bibinfo {year} {2023})}\BibitemShut {NoStop}%
\bibitem [{\citenamefont {Laughlin}(1988{\natexlab{a}})}]{Laughlin1988PRL}%
  \BibitemOpen
  \bibfield  {author} {\bibinfo {author} {\bibfnamefont {R.~B.}\ \bibnamefont {Laughlin}},\ }\bibfield  {title} {\bibinfo {title} {Superconducting ground state of noninteracting particles obeying fractional statistics},\ }\href {https://doi.org/10.1103/PhysRevLett.60.2677} {\bibfield  {journal} {\bibinfo  {journal} {Physical Review Letters}\ }\textbf {\bibinfo {volume} {60}},\ \bibinfo {pages} {2677} (\bibinfo {year} {1988}{\natexlab{a}})}\BibitemShut {NoStop}%
\bibitem [{\citenamefont {Laughlin}(1988{\natexlab{b}})}]{Laughlin1988Science}%
  \BibitemOpen
  \bibfield  {author} {\bibinfo {author} {\bibfnamefont {R.~B.}\ \bibnamefont {Laughlin}},\ }\bibfield  {title} {\bibinfo {title} {The relationship between high-temperature superconductivity and the fractional quantum hall effect},\ }\href {https://doi.org/10.1126/science.242.4878.525} {\bibfield  {journal} {\bibinfo  {journal} {Science}\ }\textbf {\bibinfo {volume} {242}},\ \bibinfo {pages} {525} (\bibinfo {year} {1988}{\natexlab{b}})}\BibitemShut {NoStop}%
\bibitem [{\citenamefont {Halperin}(1990)}]{HalperinAnyonSC1990}%
  \BibitemOpen
  \bibfield  {author} {\bibinfo {author} {\bibfnamefont {B.~I.}\ \bibnamefont {Halperin}},\ }\bibfield  {title} {\bibinfo {title} {Anyon superconductivity},\ }\href@noop {} {\bibfield  {journal} {\bibinfo  {journal} {Helv. Phys. Acta}\ }\textbf {\bibinfo {volume} {65}},\ \bibinfo {pages} {80} (\bibinfo {year} {1990})}\BibitemShut {NoStop}%
\bibitem [{\citenamefont {Fisher}\ and\ \citenamefont {Lee}(1991)}]{FisherLeeAnyonSC1991}%
  \BibitemOpen
  \bibfield  {author} {\bibinfo {author} {\bibfnamefont {M.~P.~A.}\ \bibnamefont {Fisher}}\ and\ \bibinfo {author} {\bibfnamefont {D.-H.}\ \bibnamefont {Lee}},\ }\bibfield  {title} {\bibinfo {title} {Correspondence between two‐dimensional anyons and superconductivity},\ }\href {https://doi.org/10.1103/PhysRevB.43.130} {\bibfield  {journal} {\bibinfo  {journal} {Phys. Rev. B}\ }\textbf {\bibinfo {volume} {43}},\ \bibinfo {pages} {130} (\bibinfo {year} {1991})}\BibitemShut {NoStop}%
\bibitem [{\citenamefont {Zhang}\ \emph {et~al.}(2023)\citenamefont {Zhang}, \citenamefont {Zhu},\ and\ \citenamefont {Vishwanath}}]{Zhang2023PRX}%
  \BibitemOpen
  \bibfield  {author} {\bibinfo {author} {\bibfnamefont {Y.-H.}\ \bibnamefont {Zhang}}, \bibinfo {author} {\bibfnamefont {Z.}~\bibnamefont {Zhu}},\ and\ \bibinfo {author} {\bibfnamefont {A.}~\bibnamefont {Vishwanath}},\ }\bibfield  {title} {\bibinfo {title} {\textit{XY}\textsuperscript{\*} transition and extraordinary boundary criticality from fractional exciton condensation in quantum hall bilayer},\ }\href {https://doi.org/10.1103/PhysRevX.13.031023} {\bibfield  {journal} {\bibinfo  {journal} {Physical Review X}\ }\textbf {\bibinfo {volume} {13}},\ \bibinfo {pages} {031023} (\bibinfo {year} {2023})}\BibitemShut {NoStop}%
\bibitem [{\citenamefont {Moon}\ \emph {et~al.}(1995)\citenamefont {Moon}, \citenamefont {Mori}, \citenamefont {Yang}, \citenamefont {Girvin}, \citenamefont {MacDonald}, \citenamefont {Zheng}, \citenamefont {Yoshioka},\ and\ \citenamefont {Zhang}}]{Moon1995PRB}%
  \BibitemOpen
  \bibfield  {author} {\bibinfo {author} {\bibfnamefont {K.}~\bibnamefont {Moon}}, \bibinfo {author} {\bibfnamefont {H.-C.}\ \bibnamefont {Mori}}, \bibinfo {author} {\bibfnamefont {K.}~\bibnamefont {Yang}}, \bibinfo {author} {\bibfnamefont {S.~M.}\ \bibnamefont {Girvin}}, \bibinfo {author} {\bibfnamefont {A.~H.}\ \bibnamefont {MacDonald}}, \bibinfo {author} {\bibfnamefont {L.}~\bibnamefont {Zheng}}, \bibinfo {author} {\bibfnamefont {D.}~\bibnamefont {Yoshioka}},\ and\ \bibinfo {author} {\bibfnamefont {S.-C.}\ \bibnamefont {Zhang}},\ }\bibfield  {title} {\bibinfo {title} {Spontaneous interlayer coherence in double-layer quantum hall systems: Charged vortices and kosterlitz-thouless phase transitions},\ }\href {https://doi.org/10.1103/PhysRevB.51.5138} {\bibfield  {journal} {\bibinfo  {journal} {Physical Review B}\ }\textbf {\bibinfo {volume} {51}},\ \bibinfo {pages} {5138} (\bibinfo {year} {1995})}\BibitemShut {NoStop}%
\bibitem [{\citenamefont {Spielman}\ \emph {et~al.}(2000)\citenamefont {Spielman}, \citenamefont {Eisenstein}, \citenamefont {Pfeiffer},\ and\ \citenamefont {West}}]{Spielman2000PRL}%
  \BibitemOpen
  \bibfield  {author} {\bibinfo {author} {\bibfnamefont {I.~B.}\ \bibnamefont {Spielman}}, \bibinfo {author} {\bibfnamefont {J.~P.}\ \bibnamefont {Eisenstein}}, \bibinfo {author} {\bibfnamefont {L.~N.}\ \bibnamefont {Pfeiffer}},\ and\ \bibinfo {author} {\bibfnamefont {K.~W.}\ \bibnamefont {West}},\ }\bibfield  {title} {\bibinfo {title} {Resonantly enhanced tunneling in a quantum hall ferromagnet},\ }\href {https://doi.org/10.1103/PhysRevLett.84.5808} {\bibfield  {journal} {\bibinfo  {journal} {Physical Review Letters}\ }\textbf {\bibinfo {volume} {84}},\ \bibinfo {pages} {5808} (\bibinfo {year} {2000})}\BibitemShut {NoStop}%
\bibitem [{\citenamefont {Kellogg}\ \emph {et~al.}(2004)\citenamefont {Kellogg}, \citenamefont {Eisenstein}, \citenamefont {Pfeiffer},\ and\ \citenamefont {West}}]{Kellogg2004PRL}%
  \BibitemOpen
  \bibfield  {author} {\bibinfo {author} {\bibfnamefont {M.}~\bibnamefont {Kellogg}}, \bibinfo {author} {\bibfnamefont {J.~P.}\ \bibnamefont {Eisenstein}}, \bibinfo {author} {\bibfnamefont {L.~N.}\ \bibnamefont {Pfeiffer}},\ and\ \bibinfo {author} {\bibfnamefont {K.~W.}\ \bibnamefont {West}},\ }\bibfield  {title} {\bibinfo {title} {Vanishing hall resistance at high magnetic field in a double-layer two-dimensional electron system},\ }\href {https://doi.org/10.1103/PhysRevLett.93.036801} {\bibfield  {journal} {\bibinfo  {journal} {Physical Review Letters}\ }\textbf {\bibinfo {volume} {93}},\ \bibinfo {pages} {036801} (\bibinfo {year} {2004})}\BibitemShut {NoStop}%
\bibitem [{\citenamefont {Tutuc}\ \emph {et~al.}(2004)\citenamefont {Tutuc}, \citenamefont {Shayegan},\ and\ \citenamefont {Huse}}]{Tutuc2004PRL}%
  \BibitemOpen
  \bibfield  {author} {\bibinfo {author} {\bibfnamefont {E.}~\bibnamefont {Tutuc}}, \bibinfo {author} {\bibfnamefont {M.}~\bibnamefont {Shayegan}},\ and\ \bibinfo {author} {\bibfnamefont {D.~A.}\ \bibnamefont {Huse}},\ }\bibfield  {title} {\bibinfo {title} {Counterflow resistivity in strongly correlated gaas bilayers},\ }\href {https://doi.org/10.1103/PhysRevLett.93.036802} {\bibfield  {journal} {\bibinfo  {journal} {Physical Review Letters}\ }\textbf {\bibinfo {volume} {93}},\ \bibinfo {pages} {036802} (\bibinfo {year} {2004})}\BibitemShut {NoStop}%
\bibitem [{\citenamefont {Nandi}\ \emph {et~al.}(2012)\citenamefont {Nandi}, \citenamefont {Finck}, \citenamefont {Eisenstein}, \citenamefont {Pfeiffer},\ and\ \citenamefont {West}}]{Nandi2012Nature}%
  \BibitemOpen
  \bibfield  {author} {\bibinfo {author} {\bibfnamefont {D.}~\bibnamefont {Nandi}}, \bibinfo {author} {\bibfnamefont {A.~D.~K.}\ \bibnamefont {Finck}}, \bibinfo {author} {\bibfnamefont {J.~P.}\ \bibnamefont {Eisenstein}}, \bibinfo {author} {\bibfnamefont {L.~N.}\ \bibnamefont {Pfeiffer}},\ and\ \bibinfo {author} {\bibfnamefont {K.~W.}\ \bibnamefont {West}},\ }\bibfield  {title} {\bibinfo {title} {Exciton condensation and perfect coulomb drag},\ }\href {https://doi.org/10.1038/nature11302} {\bibfield  {journal} {\bibinfo  {journal} {Nature}\ }\textbf {\bibinfo {volume} {488}},\ \bibinfo {pages} {481} (\bibinfo {year} {2012})}\BibitemShut {NoStop}%
\bibitem [{\citenamefont {Li}\ \emph {et~al.}(2017)\citenamefont {Li}, \citenamefont {Taniguchi}, \citenamefont {Watanabe}, \citenamefont {Hone}, \citenamefont {Dean},\ and\ \citenamefont {Young}}]{Li2017NatPhys}%
  \BibitemOpen
  \bibfield  {author} {\bibinfo {author} {\bibfnamefont {J.~I.~A.}\ \bibnamefont {Li}}, \bibinfo {author} {\bibfnamefont {T.}~\bibnamefont {Taniguchi}}, \bibinfo {author} {\bibfnamefont {K.}~\bibnamefont {Watanabe}}, \bibinfo {author} {\bibfnamefont {J.}~\bibnamefont {Hone}}, \bibinfo {author} {\bibfnamefont {C.~R.}\ \bibnamefont {Dean}},\ and\ \bibinfo {author} {\bibfnamefont {A.~F.}\ \bibnamefont {Young}},\ }\bibfield  {title} {\bibinfo {title} {Excitonic superfluid phase in double bilayer graphene},\ }\href {https://doi.org/10.1038/nphys4140} {\bibfield  {journal} {\bibinfo  {journal} {Nature Physics}\ }\textbf {\bibinfo {volume} {13}},\ \bibinfo {pages} {751} (\bibinfo {year} {2017})}\BibitemShut {NoStop}%
\bibitem [{\citenamefont {Liu}\ \emph {et~al.}(2017)\citenamefont {Liu}, \citenamefont {Watanabe}, \citenamefont {Taniguchi}, \citenamefont {Halperin},\ and\ \citenamefont {Kim}}]{Liu2017NatPhys}%
  \BibitemOpen
  \bibfield  {author} {\bibinfo {author} {\bibfnamefont {X.}~\bibnamefont {Liu}}, \bibinfo {author} {\bibfnamefont {K.}~\bibnamefont {Watanabe}}, \bibinfo {author} {\bibfnamefont {T.}~\bibnamefont {Taniguchi}}, \bibinfo {author} {\bibfnamefont {B.~I.}\ \bibnamefont {Halperin}},\ and\ \bibinfo {author} {\bibfnamefont {P.}~\bibnamefont {Kim}},\ }\bibfield  {title} {\bibinfo {title} {Quantum hall drag of exciton condensate in graphene},\ }\href {https://doi.org/10.1038/nphys4144} {\bibfield  {journal} {\bibinfo  {journal} {Nature Physics}\ }\textbf {\bibinfo {volume} {13}},\ \bibinfo {pages} {746} (\bibinfo {year} {2017})}\BibitemShut {NoStop}%
\bibitem [{\citenamefont {Liu}\ \emph {et~al.}(2022)\citenamefont {Liu}, \citenamefont {Zeng}, \citenamefont {Li},\ and\ \citenamefont {Dean}}]{Liu2022Science}%
  \BibitemOpen
  \bibfield  {author} {\bibinfo {author} {\bibfnamefont {X.}~\bibnamefont {Liu}}, \bibinfo {author} {\bibfnamefont {Y.}~\bibnamefont {Zeng}}, \bibinfo {author} {\bibfnamefont {J.~I.~A.}\ \bibnamefont {Li}},\ and\ \bibinfo {author} {\bibfnamefont {C.~R.}\ \bibnamefont {Dean}},\ }\bibfield  {title} {\bibinfo {title} {Crossover between strongly and weakly coupled exciton superfluids},\ }\href {https://doi.org/10.1126/science.abg1110} {\bibfield  {journal} {\bibinfo  {journal} {Science}\ }\textbf {\bibinfo {volume} {375}},\ \bibinfo {pages} {205} (\bibinfo {year} {2022})}\BibitemShut {NoStop}%
\bibitem [{\citenamefont {Jo}\ \emph {et~al.}(1992)\citenamefont {Jo}, \citenamefont {Sachrajda},\ and\ \citenamefont {Williams}}]{Jo1992PRB}%
  \BibitemOpen
  \bibfield  {author} {\bibinfo {author} {\bibfnamefont {J.}~\bibnamefont {Jo}}, \bibinfo {author} {\bibfnamefont {A.~S.}\ \bibnamefont {Sachrajda}},\ and\ \bibinfo {author} {\bibfnamefont {R.~L.}\ \bibnamefont {Williams}},\ }\bibfield  {title} {\bibinfo {title} {Quantum hall effect in a triple-layer electron system},\ }\href {https://doi.org/10.1103/PhysRevB.46.9776} {\bibfield  {journal} {\bibinfo  {journal} {Physical Review B}\ }\textbf {\bibinfo {volume} {46}},\ \bibinfo {pages} {9776} (\bibinfo {year} {1992})}\BibitemShut {NoStop}%
\bibitem [{\citenamefont {Zeng}(2021)}]{Zeng2021Thesis}%
  \BibitemOpen
  \bibfield  {author} {\bibinfo {author} {\bibfnamefont {Y.}~\bibnamefont {Zeng}},\ }\emph {\bibinfo {title} {Study of Two-Dimensional Correlated Quantum Fluid in Multi-layer Graphene System}},\ \href@noop {} {\bibinfo {type} {Ph.d. thesis}},\ \bibinfo  {school} {Columbia University}, \bibinfo {address} {New York, NY, USA} (\bibinfo {year} {2021})\BibitemShut {NoStop}%
\bibitem [{\citenamefont {Almoalem}\ \emph {et~al.}(2022)\citenamefont {Almoalem}, \citenamefont {Feldman}, \citenamefont {Mangel}, \citenamefont {Bulk}, \citenamefont {Ruhman}, \citenamefont {Moshe},\ and\ \citenamefont {Kanigel}}]{Almoalem2022NatPhys}%
  \BibitemOpen
  \bibfield  {author} {\bibinfo {author} {\bibfnamefont {A.}~\bibnamefont {Almoalem}}, \bibinfo {author} {\bibfnamefont {I.}~\bibnamefont {Feldman}}, \bibinfo {author} {\bibfnamefont {I.}~\bibnamefont {Mangel}}, \bibinfo {author} {\bibfnamefont {Y.}~\bibnamefont {Bulk}}, \bibinfo {author} {\bibfnamefont {J.}~\bibnamefont {Ruhman}}, \bibinfo {author} {\bibfnamefont {M.}~\bibnamefont {Moshe}},\ and\ \bibinfo {author} {\bibfnamefont {A.}~\bibnamefont {Kanigel}},\ }\bibfield  {title} {\bibinfo {title} {State with spontaneously broken time-reversal symmetry above the superconducting phase transition},\ }\href {https://doi.org/10.1038/s41567-022-01709-9} {\bibfield  {journal} {\bibinfo  {journal} {Nature Physics}\ }\textbf {\bibinfo {volume} {18}},\ \bibinfo {pages} {869} (\bibinfo {year} {2022})}\BibitemShut {NoStop}%
\bibitem [{\citenamefont {Iyuha}\ \emph {et~al.}(2024)\citenamefont {Iyuha}, \citenamefont {Almoalem}, \citenamefont {Kanigel},\ and\ \citenamefont {Ruhman}}]{Iyuha2024LP}%
  \BibitemOpen
  \bibfield  {author} {\bibinfo {author} {\bibfnamefont {Y.}~\bibnamefont {Iyuha}}, \bibinfo {author} {\bibfnamefont {A.}~\bibnamefont {Almoalem}}, \bibinfo {author} {\bibfnamefont {A.}~\bibnamefont {Kanigel}},\ and\ \bibinfo {author} {\bibfnamefont {J.}~\bibnamefont {Ruhman}},\ }\bibfield  {title} {\bibinfo {title} {Observation of $\pi$-shift little–parks oscillations in 4hb-tas$_2$ single-crystal rings},\ }\href {https://doi.org/10.1038/s41567-024-02099-2} {\bibfield  {journal} {\bibinfo  {journal} {Nature Physics}\ }\textbf {\bibinfo {volume} {20}},\ \bibinfo {pages} {1254} (\bibinfo {year} {2024})}\BibitemShut {NoStop}%
\bibitem [{\citenamefont {Iguchi}\ \emph {et~al.}(2023)\citenamefont {Iguchi}, \citenamefont {Shi}, \citenamefont {Kihou}, \citenamefont {Lee}, \citenamefont {Huang}, \citenamefont {Hassinger}, \citenamefont {Wang}, \citenamefont {Kasahara}, \citenamefont {Mazin}, \citenamefont {Wang}, \citenamefont {Maeno}, \citenamefont {Mackenzie},\ and\ \citenamefont {Davis}}]{Iguchi2023Science}%
  \BibitemOpen
  \bibfield  {author} {\bibinfo {author} {\bibfnamefont {Y.}~\bibnamefont {Iguchi}}, \bibinfo {author} {\bibfnamefont {R.}~\bibnamefont {Shi}}, \bibinfo {author} {\bibfnamefont {K.}~\bibnamefont {Kihou}}, \bibinfo {author} {\bibfnamefont {C.-H.}\ \bibnamefont {Lee}}, \bibinfo {author} {\bibfnamefont {C.-L.}\ \bibnamefont {Huang}}, \bibinfo {author} {\bibfnamefont {E.}~\bibnamefont {Hassinger}}, \bibinfo {author} {\bibfnamefont {Z.}~\bibnamefont {Wang}}, \bibinfo {author} {\bibfnamefont {S.}~\bibnamefont {Kasahara}}, \bibinfo {author} {\bibfnamefont {I.~I.}\ \bibnamefont {Mazin}}, \bibinfo {author} {\bibfnamefont {Y.}~\bibnamefont {Wang}}, \bibinfo {author} {\bibfnamefont {Y.}~\bibnamefont {Maeno}}, \bibinfo {author} {\bibfnamefont {A.~P.}\ \bibnamefont {Mackenzie}},\ and\ \bibinfo {author} {\bibfnamefont {J.~C.~S.}\ \bibnamefont {Davis}},\ }\bibfield  {title} {\bibinfo {title} {Superconducting vortices carrying a temperature-dependent fraction of the flux quantum},\ }\href
  {https://doi.org/10.1126/science.abp9979} {\bibfield  {journal} {\bibinfo  {journal} {Science}\ }\textbf {\bibinfo {volume} {380}},\ \bibinfo {pages} {eabp9979} (\bibinfo {year} {2023})}\BibitemShut {NoStop}%
\bibitem [{\citenamefont {Ge}\ \emph {et~al.}(2024)\citenamefont {Ge}, \citenamefont {Wang}, \citenamefont {Xing}, \citenamefont {Yin}, \citenamefont {Wang}, \citenamefont {Shen}, \citenamefont {Lei}, \citenamefont {Wang},\ and\ \citenamefont {Wang}}]{Ge2024PRX}%
  \BibitemOpen
  \bibfield  {author} {\bibinfo {author} {\bibfnamefont {J.}~\bibnamefont {Ge}}, \bibinfo {author} {\bibfnamefont {P.}~\bibnamefont {Wang}}, \bibinfo {author} {\bibfnamefont {Y.}~\bibnamefont {Xing}}, \bibinfo {author} {\bibfnamefont {Q.}~\bibnamefont {Yin}}, \bibinfo {author} {\bibfnamefont {A.}~\bibnamefont {Wang}}, \bibinfo {author} {\bibfnamefont {J.}~\bibnamefont {Shen}}, \bibinfo {author} {\bibfnamefont {H.}~\bibnamefont {Lei}}, \bibinfo {author} {\bibfnamefont {Z.}~\bibnamefont {Wang}},\ and\ \bibinfo {author} {\bibfnamefont {J.}~\bibnamefont {Wang}},\ }\bibfield  {title} {\bibinfo {title} {Charge-$4e$ and charge-$6e$ flux quantization and higher charge superconductivity in kagome superconductor ring devices},\ }\href {https://doi.org/10.1103/PhysRevX.14.021025} {\bibfield  {journal} {\bibinfo  {journal} {Physical Review X}\ }\textbf {\bibinfo {volume} {14}},\ \bibinfo {pages} {021025} (\bibinfo {year} {2024})}\BibitemShut {NoStop}%
\bibitem [{\citenamefont {Zaletel}\ and\ \citenamefont {Mong}(2012)}]{ExactMPS}%
  \BibitemOpen
  \bibfield  {author} {\bibinfo {author} {\bibfnamefont {M.~P.}\ \bibnamefont {Zaletel}}\ and\ \bibinfo {author} {\bibfnamefont {R.~S.~K.}\ \bibnamefont {Mong}},\ }\bibfield  {title} {\bibinfo {title} {Exact matrix product states for quantum hall wave functions},\ }\href {https://doi.org/10.1103/PhysRevB.86.245305} {\bibfield  {journal} {\bibinfo  {journal} {Phys. Rev. B}\ }\textbf {\bibinfo {volume} {86}},\ \bibinfo {pages} {245305} (\bibinfo {year} {2012})}\BibitemShut {NoStop}%
\bibitem [{\citenamefont {Zaletel}\ \emph {et~al.}(2013)\citenamefont {Zaletel}, \citenamefont {Mong},\ and\ \citenamefont {Pollmann}}]{TopoChara}%
  \BibitemOpen
  \bibfield  {author} {\bibinfo {author} {\bibfnamefont {M.~P.}\ \bibnamefont {Zaletel}}, \bibinfo {author} {\bibfnamefont {R.~S.~K.}\ \bibnamefont {Mong}},\ and\ \bibinfo {author} {\bibfnamefont {F.}~\bibnamefont {Pollmann}},\ }\bibfield  {title} {\bibinfo {title} {Topological characterization of fractional quantum hall ground states from microscopic hamiltonians},\ }\href {https://doi.org/10.1103/PhysRevLett.110.236801} {\bibfield  {journal} {\bibinfo  {journal} {Phys. Rev. Lett.}\ }\textbf {\bibinfo {volume} {110}},\ \bibinfo {pages} {236801} (\bibinfo {year} {2013})}\BibitemShut {NoStop}%
\bibitem [{\citenamefont {Liu}\ \emph {et~al.}(2021)\citenamefont {Liu}, \citenamefont {Li},\ and\ \citenamefont {Dean}}]{Liu2021PRL}%
  \BibitemOpen
  \bibfield  {author} {\bibinfo {author} {\bibfnamefont {X.}~\bibnamefont {Liu}}, \bibinfo {author} {\bibfnamefont {J.~I.~A.}\ \bibnamefont {Li}},\ and\ \bibinfo {author} {\bibfnamefont {C.~R.}\ \bibnamefont {Dean}},\ }\bibfield  {title} {\bibinfo {title} {Frictional magneto–coulomb drag in graphene double layers},\ }\href {https://doi.org/10.1103/PhysRevLett.127.047702} {\bibfield  {journal} {\bibinfo  {journal} {Physical Review Letters}\ }\textbf {\bibinfo {volume} {127}},\ \bibinfo {pages} {047702} (\bibinfo {year} {2021})}\BibitemShut {NoStop}%
\bibitem [{\citenamefont {Zibrov}\ \emph {et~al.}(2017)\citenamefont {Zibrov}, \citenamefont {Kometter}, \citenamefont {Zhou}, \citenamefont {Taniguchi}, \citenamefont {Watanabe}, \citenamefont {Zaletel},\ and\ \citenamefont {Young}}]{Zibrov2017Nature}%
  \BibitemOpen
  \bibfield  {author} {\bibinfo {author} {\bibfnamefont {A.~A.}\ \bibnamefont {Zibrov}}, \bibinfo {author} {\bibfnamefont {C.}~\bibnamefont {Kometter}}, \bibinfo {author} {\bibfnamefont {H.}~\bibnamefont {Zhou}}, \bibinfo {author} {\bibfnamefont {T.}~\bibnamefont {Taniguchi}}, \bibinfo {author} {\bibfnamefont {K.}~\bibnamefont {Watanabe}}, \bibinfo {author} {\bibfnamefont {M.~P.}\ \bibnamefont {Zaletel}},\ and\ \bibinfo {author} {\bibfnamefont {A.~F.}\ \bibnamefont {Young}},\ }\bibfield  {title} {\bibinfo {title} {Tunable interacting composite fermion phases in a half‐filled bilayer‐graphene landau level},\ }\href {https://doi.org/10.1038/nature23893} {\bibfield  {journal} {\bibinfo  {journal} {Nature}\ }\textbf {\bibinfo {volume} {549}},\ \bibinfo {pages} {360} (\bibinfo {year} {2017})}\BibitemShut {NoStop}%
\bibitem [{\citenamefont {Hartman}\ \emph {et~al.}(2018)\citenamefont {Hartman}, \citenamefont {Erlandsen},\ and\ \citenamefont {Sudb{\o}}}]{Hartman2018PRB}%
  \BibitemOpen
  \bibfield  {author} {\bibinfo {author} {\bibfnamefont {S.}~\bibnamefont {Hartman}}, \bibinfo {author} {\bibfnamefont {E.}~\bibnamefont {Erlandsen}},\ and\ \bibinfo {author} {\bibfnamefont {A.}~\bibnamefont {Sudb{\o}}},\ }\bibfield  {title} {\bibinfo {title} {Superfluid drag in multicomponent bose--einstein condensates on a square optical lattice},\ }\href {https://doi.org/10.1103/PhysRevB.98.024512} {\bibfield  {journal} {\bibinfo  {journal} {Physical Review B}\ }\textbf {\bibinfo {volume} {98}},\ \bibinfo {pages} {024512} (\bibinfo {year} {2018})}\BibitemShut {NoStop}%
\bibitem [{\citenamefont {Wang}\ \emph {et~al.}(2024)\citenamefont {Wang}, \citenamefont {Fan}, \citenamefont {Dai},\ and\ \citenamefont {Zaletel}}]{Wang2024JJ}%
  \BibitemOpen
  \bibfield  {author} {\bibinfo {author} {\bibfnamefont {T.}~\bibnamefont {Wang}}, \bibinfo {author} {\bibfnamefont {R.}~\bibnamefont {Fan}}, \bibinfo {author} {\bibfnamefont {Z.}~\bibnamefont {Dai}},\ and\ \bibinfo {author} {\bibfnamefont {M.~P.}\ \bibnamefont {Zaletel}},\ }\bibfield  {title} {\bibinfo {title} {Designing exciton-condensate josephson junctions in quantum hall heterostructures},\ }\href@noop {} {\bibfield  {journal} {\bibinfo  {journal} {arXiv}\ } (\bibinfo {year} {2024})},\ \bibinfo {note} {cond-mat.mes-hall},\ \Eprint {https://arxiv.org/abs/2409.19059} {2409.19059} \BibitemShut {NoStop}%
\end{thebibliography}%

\end{document}